\documentclass[12pt,american]{revtex4}
\usepackage[T1]{fontenc}
\usepackage[latin1]{inputenc}
\usepackage{amsmath}
\usepackage{graphicx}
\usepackage{amssymb}

\usepackage{graphicx}

\usepackage{babel}

\begin{document}

\title{Particle creation in a tunneling universe}

\author{Jooyoo Hong$^{1,2}$, Alexander Vilenkin$^{1}$, and Serge Winitzki$^{1}$}

\affiliation{$^{1}$Institute of Cosmology, Department of Physics and Astronomy,
Tufts University, Medford, MA 02155, USA}

\affiliation{$^{2}$ Department of Physics, Hanyang University at Ansan, Ansan,
Kyunggi-do, 425-791 Korea}

\begin{abstract}
An expanding closed universe filled with radiation can either recollapse
or tunnel to the regime of unbounded expansion, if the cosmological
constant is nonzero. We re-examine the question of particle creation
during tunneling, with the purpose of resolving a long-standing controversy.
Using a perturbative superspace model with a conformally coupled massless
scalar field, which is known to give no particle production, we explicitly
show that the breakdown of the semiclassical approximation and the
{}``catastrophic particle production'' claimed earlier in the literature
are due to an inappropriate choice of the initial quantum state prior
to the tunneling. 
\end{abstract}
\maketitle

\section{Introduction}

The tunneling approach to quantum cosmology proposes to view the creation
of the Universe as a quantum tunneling event \cite{AV86,R84,ZelStar}.
A semiclassical picture based on the Wheeler-DeWitt equation describes
tunneling from a state of vanishing size ({}``tunneling from nothing'')
to a closed inflating universe. Alternative proposals have been put
forward by Hartle and Hawking \cite{HH} and Linde \cite{Linde84}.
For a recent critical review see, e.g., \cite{AV02}. In the present
paper, we shall focus on the tunneling picture.

The process of tunneling from nothing can be thought of as a limit
of tunneling from a closed recollapsing universe of very small but
nonzero size ({}``tunneling from something''). Consider a closed
FRW universe filled with a vacuum of constant energy density and some
matter or radiation. Classically, such a universe has two possible
types of evolution. It can expand to a maximum radius and recollapse,
or it can contract from an infinite size, bounce at a minimum radius
and then reexpand. But in quantum cosmology there is yet another possibility.
Instead of recollapsing, the universe can tunnel through a potential
barrier to the regime of unbounded expansion. The semiclassical tunneling
probability does not vanish in the limit when the maximum size of
the initial universe shrinks to zero. The corresponding wave function
describes quantum nucleation of an inflating universe from nothing.

The question of the particle content in the universe created by quantum
tunneling has been studied in the literature by including the inhomogeneous
modes of quantum fields in the wave function. Conflicting claims of
excessive particle production during tunneling \cite{R84,LRR02},
on the one hand, and of essentially no particle content in the inflating
universe \cite{V88,VV88,GV97}, on the other hand, have been advanced.
In this paper we intend to resolve this controversy and explain the
origin of the conflicting results.

Rubakov \cite{R84} (see also \cite{LRR02}) considered scalar particle
production in the process of {}``tunneling from something''. He
found that, for a generic initial state of the universe, the wave
function is significantly affected by the apparent growth of the excitations
of the scalar field during tunneling. Rubakov interpreted this result
as an excessive particle production during tunneling and a breakdown
of the semiclassical evolution. However, we shall argue that this
result is a consequence of an inappropriate choice of the quantum
state of the universe and that in fact there is only a finite, if
any, particle production during tunneling.

Our main points can be summarized as follows. An arbitrary initial
state of the recollapsing universe can be interpreted as a superposition
of semiclassical geometries with certain amplitudes, each branch being
a classical universe with a certain quantum state of the scalar field.
In a generic superposition, essentially all excited states of the
scalar field will be represented, perhaps with small amplitudes. In
this case, some of the semiclassical branches will have such a large
energy due to the excitations of the scalar field that no tunneling
will take place: in these branches, the initial universe will not
recollapse but will continue expanding. Non-tunneling branches give
a potentially larger contribution to the wave function at large scale
factors because they are not exponentially suppressed, compared with
the tunneling branches. If the amplitudes of such high-energy branches
in a given superposition are sufficiently large, they will give a
dominant contribution to the wave function of the inflating universe,
while the contribution of the tunneling branches will be negligible.

If one considers the tunneling from a recollapsing universe of nonzero
size ({}``tunneling from something''), then one is free to choose
the initial quantum state of that universe. We shall show that a consistent
semiclassical picture of the tunneling universe can only be obtained
with an appropriate choice of the initial quantum state of the recollapsing
universe. This state should be such that both the recollapsing and
the inflating universe belong to the same semiclassical branch of
the wave function. Rubakov \emph{et al.}~\cite{R84,LRR02} have chosen
a quantum state that does not satisfy this condition. On the other
hand, Refs.~\cite{V88,VV88} have considered {}``tunneling from
nothing'' (the limit of zero size of the recollapsing universe) using
a quantum state that describes a single underlying semiclassical geometry.
In the present paper we shall explicitly construct such quantum states
for the case of {}``tunneling from something''. The difference in
the choice of the quantum state is the first cause of the discrepancy
in the cited papers.

The second problem with the results of Rubakov \emph{et al.~}is their
particle interpretation of the wave function. A particle interpretation
of a quantum field theory requires a fixed, classical background metric.
With a generic choice of the quantum state of the universe, the Wheeler-DeWitt
wave function will describe a quantum superposition of different geometries,
rather than a single classical geometry. In this case, the interpretation
of the Wheeler-DeWitt wave function that Rubakov \emph{et al.}~used
to obtain the particle content (the formalism of the instantaneous
Hamiltonian diagonalization) is not justified. If the particle content
is inferred from the wave function as if there exists a unique underlying
semiclassical spacetime, then one is lead to erroneous conclusions
about the breakdown of the tunneling process and about the excessive
particle production.

The paper is organized as follows. In Sec.~2, we review the {}``perturbative
superspace'' approach to the Wheeler-DeWitt equation. We demonstrate
that a Gaussian ansatz for the WKB wave function employed in Refs.~\cite{V88,VV88}
corresponds to an {}``instantaneous squeezed state'' in the formalism
of Rubakov \emph{et al.} In this sense we find a formal agreement
between these calculations. In Sec.~3, we consider a massless conformally
coupled scalar field, in which case the Wheeler-DeWitt equation is
separable and it is well known that there is no particle production
\cite{Parker}. Assuming a squeezed initial quantum state for the
scalar field, we show by an explicit calculation that the wave function
becomes dominated by high-energy states far enough under the barrier,
and that the interpretation of Rubakov \emph{et al.~}would indicate
a {}``catastrophic particle production'' during tunneling. Details
of the calculation are given in Appendix A. In Sec.~4, we discuss
what we believe to be the correct physical interpretation of the results.
The more complicated case of a massive scalar field will be presented
in the companion paper \cite{JVW2}.

\section{Semiclassical perturbative superspace}

We consider a homogeneous (closed) FRW universe with a conformally
coupled scalar field $\phi $. The metric is homogeneous on 3-spheres,\begin{equation}
ds^{2}=dt^{2}-a^{2}\left(t\right)d\Omega _{3}^{2},\end{equation}
 where $a\left(t\right)$ is the scale factor. The scalar field $\phi $
is not homogeneous and may be expanded in 3-spherical harmonics,\begin{equation}
\phi \left(x,t\right)=\frac{\pi \sqrt{2}}{a\left(t\right)}\sum _{n,l,p}\chi _{nlp}\left(t\right)Q_{lp}^{n}\left(\mathbf{x}\right).\end{equation}
 Below, only the index $n=1,2,...$ will enter the equations, and
we shall suppress the indices $l$, $p$ of the modes $\chi _{nlp}\left(t\right)$.
The summation over degenerate indices $l$, $p$ spans $l=0$, ...,
$n-1$ and $p=-l$, ..., $l$ and introduces an extra factor $n^{2}$
which we shall insert in explicit calculations below.

The model is described by the classical action\begin{equation}
\int d^{4}x\sqrt{-g}\left\{ \frac{1}{16\pi }R+\frac{1}{2}\left(\partial _{\mu }\phi \right)^{2}-\frac{3}{8\pi }H^{2}-\frac{m^{2}\phi ^{2}}{2}-\frac{1}{12}R\phi ^{2}\right\} .\end{equation}
 Here, the parameter $H$ represents the vacuum energy (the cosmological
constant), $R$ is the scalar curvature, $m$ is the mass of the scalar
field, and we are using Planck units, $G=\hbar =c=1$. In the Schrödinger
picture of the {}``perturbative superspace'' approach to quantum
gravity \cite{HH85}, the wave function of the universe $\Psi \left(a,\left\{ \chi _{n}\right\} \right)$
after appropriate rescalings of the parameters (see \cite{V88} for
more details) satisfies the Wheeler-DeWitt equation\begin{equation}
\left[\hbar ^{2}\frac{\partial ^{2}}{\partial a^{2}}-V\left(a\right)-\sum _{n}\mathcal{H}_{n}\right]\Psi \left(a,\left\{ \chi _{n}\right\} \right)=0.\label{eq:wdw}\end{equation}
 Here, the scalar field Hamiltonian for the $n$-th mode is\begin{equation}
\mathcal{H}_{n}\equiv \hbar ^{2}\frac{\partial ^{2}}{\partial \chi _{n}^{2}}-\left(n^{2}+m^{2}a^{2}\right)\chi _{n}^{2},\label{eq:chinHn}\end{equation}
 and \begin{equation}
V\left(a\right)\equiv a^{2}-H^{2}a^{4}.\end{equation}
 We have written out the Planck constant $\hbar $ in Eq.~(\ref{eq:wdw})
to make the WKB approximation more explicit below. {[}The factor $\hbar $
is merely a formal bookkeeping parameter since $\hbar \equiv 1$ in
Planck units.{]} The Hamiltonian of Eq.~(\ref{eq:chinHn}) describes
a harmonic oscillator with an $a$-dependent frequency\begin{equation}
\omega _{n}=\omega _{n}\left(a\right)\equiv \sqrt{n^{2}+m^{2}a^{2}}.\label{eq:omega-def}\end{equation}
 As in Ref.~\cite{R84}, we shall include in addition to the scalar
field a homogeneous radiation component with energy density \begin{equation}
\rho _{r}=a^{-4}\varepsilon _{r},\end{equation}
 where $\varepsilon _{r}$ is a constant parameter. This amounts to
replacing\begin{equation}
H^{2}\rightarrow H^{2}+\frac{\varepsilon _{r}}{a^{4}}\end{equation}
 and therefore Eq.~(\ref{eq:wdw}) still holds with\begin{equation}
V\left(a\right)=a^{2}-H^{2}a^{4}-\varepsilon _{r}.\end{equation}

\begin{figure}[htbp]
\begin{center}\includegraphics[  width=3in,
  keepaspectratio]{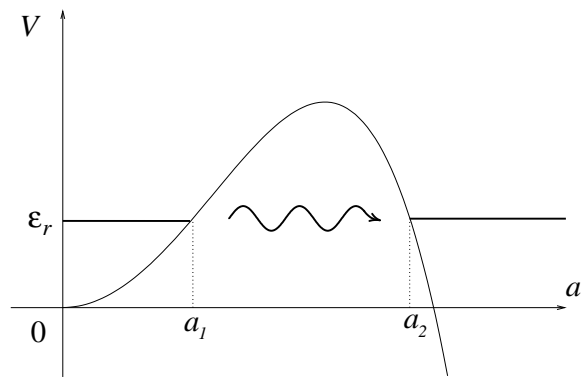}\end{center}

\caption{{}``Tunneling from something'': creation of the inflating universe
by tunneling.\label{cap:fig1a}}
\end{figure}

The physical picture of the universe in this model is illustrated
in Fig.~\ref{cap:fig1a}. The Wheeler-DeWitt equation in the $\left(a,\chi _{n}\right)$
space is formally similar to a stationary Schrödinger equation for
a quantum-mechanical particle in a two-dimensional potential. A small
closed universe filled with radiation of energy density $\varepsilon _{r}a^{-4}$
is expanding and recollapsing in the neighborhood of $a=0$, and an
inflating universe with scale factor $a_{2}$ is created by tunneling
through the potential barrier. The maximum scale factor $a_{1}$ of
the recollapsing universe is determined by the energy parameter $\varepsilon _{r}$.

A typical solution of the Wheeler-DeWitt equation that represents
the creation of an inflating universe by tunneling is qualitatively
analogous to the quantum-mechanical tunneling wave function with the
above potential. In the regions outside the barrier, the wave function
is oscillatory (in the $a$ direction) and generally has the form
of a linear combination of incoming and outgoing waves,\begin{equation}
\Psi \left(a,\left\{ \chi _{n}\right\} \right)=f_{1}\left(\left\{ \chi _{n}\right\} \right)e^{iS_{1}\left(a\right)}+f_{2}\left(\left\{ \chi _{n}\right\} \right)e^{-iS_{2}\left(a\right)}.\end{equation}
 The tunneling boundary condition specifies that the wave function
should contain only the outgoing wave in the domain $a>a_{2}$. In
this domain, we can interpret the wave function as describing a classical
spacetime if we introduce a semiclassical time variable which is a
function of $a$ \cite{LR79,Banks}. A customary ansatz is to introduce
the conformal time by\begin{equation}
d\tau =\frac{da}{\sqrt{-V\left(a\right)}}.\end{equation}

In the underbarrier region, $a_{1}<a<a_{2}$, the wave function is
a linear combination of growing and decaying real exponentials; the
conformal time becomes imaginary. The underbarrier region $a_{1}<a<a_{2}$
does not correspond to a classical spacetime.

\subsection{The Gaussian approximation}

To obtain an approximate solution, one can substitute the following
WKB-motivated Gaussian ansatz into Eq.~(\ref{eq:wdw}),\begin{equation}
\Psi \left(a,\left\{ \chi _{n}\right\} \right)=\exp \left[-\frac{1}{\hbar }S\left(a\right)-\frac{1}{2\hbar }\sum _{n}S_{n}\left(a\right)\chi _{n}^{2}\right]\label{eq:wfansatz}\end{equation}
 and neglect the terms of order $O\left(\chi _{n}^{4}\right)$. (This
ansatz was introduced in Ref.~\cite{BBW73} in the context of tunneling
in two dimensions.) Equation (\ref{eq:wfansatz}) approximates the
wave function only near $\chi _{n}=0$, where higher powers of $\chi _{n}$
are negligible. We shall refer to the wave function of Eq.~(\ref{eq:wfansatz})
simply as the Gaussian solution. In the companion paper \cite{JVW2}
we shall give a detailed analysis of the applicability of this approximation.

The functions $S\left(a\right)$, $S_{n}\left(a\right)$ in the ansatz
are to be found from the equations\begin{eqnarray}
\left(S^{\prime }\right)^{2}-V\left(a\right)-\hbar S^{\prime \prime }+\hbar \sum _{n}S_{n} & = & 0,\label{eq:S0equ}\\
S^{\prime }S_{n}^{\prime }-S_{n}^{2}+n^{2}+m^{2}a^{2}-\frac{\hbar }{2}S_{n}^{\prime \prime } & = & 0.\label{eq:Snequ}
\end{eqnarray}
 The WKB approximation consists of neglecting the terms of order $O\left(\hbar \right)$
in Eqs.~(\ref{eq:S0equ})-(\ref{eq:Snequ}). {[}Note that the last
term in Eq.~(\ref{eq:S0equ}) can be interpreted as the backreaction
of the scalar field excitations on the background metric.{]}

In the underbarrier region $a_{1}<a<a_{2}$, we introduce the Euclidean
conformal time variable $\tau $ by\begin{equation}
\tau \left(a\right)\equiv \int _{a_{1}}^{a}\frac{da}{\sqrt{V\left(a\right)}}.\label{eq:taudef}\end{equation}
 {[}A straightforward analytic continuation of this definition is
to be used in the classically allowed regions.{]} Then we obtain the
equations\begin{eqnarray}
S\left(a\right) & = & \pm \int da\sqrt{V\left(a\right)},\label{eq:S0equWKB}\\
\frac{dS_{n}}{d\tau } & = & S_{n}^{2}-\omega _{n}^{2}.\label{eq:SnequWKB}
\end{eqnarray}

A perhaps unexpected property of the Gaussian solution is that the
real part of the function $S_{n}\left(a\right)$ obtained from Eq.~(\ref{eq:SnequWKB})
may become negative at some values of $a$ under the barrier, making
the Gaussian wave function grow at large $\chi _{n}$. Whether or
not this happens is determined by the initial value $S_{n}\left(a_{1}\right)$
at the first turning point. In previous work \cite{VV88}, the condition\begin{equation}
Re\, S_{n}>0\label{eq:Sn-pos}\end{equation}
 was motivated by the requirement that the wave function be finite
at large $\chi _{n}$. This condition has been used in Refs.~\cite{VV88,GV97}
and yielded physically reasonable results. However, it should be noted
that, strictly speaking, the occurrence of $Re\, S_{n}\leq 0$ is
not necessarily problematic, since the Gaussian approximation should
only be applicable at small $\chi _{n}$. We shall see below that
the onset of the {}``catastrophic particle production'' as claimed
by Rubakov \emph{et al.}~is directly related to the change of sign
of $Re\, S_{n}\left(a\right)$. Our considerations will clarify the
physical interpretation of this phenomenon and justify the condition
of Eq.~(\ref{eq:Sn-pos}).

\subsection{The instantaneous diagonalization picture}

Rubakov \emph{et al.~}\cite{R84,LRR02} have used the method of instantaneous
Hamiltonian diagonalization to solve Eq.~(\ref{eq:wdw}). This is
equivalent to expanding the wave function $\Psi \left(a,\left\{ \chi _{n}\right\} \right)$
in the eigenstates $\psi _{k}^{(n)}\left(a,\chi _{n}\right)$ of the
$a$-dependent Hamiltonian of Eq.~(\ref{eq:chinHn}). The normalized
instantaneous eigenstates can be taken as\begin{equation}
\psi _{k}^{(n)}\left(a,\chi _{n}\right)=\frac{\left(\frac{\omega _{n}}{\pi \hbar }\right)^{\frac{1}{4}}}{\sqrt{2^{k}k!}}H_{k}\left(\chi _{n}\sqrt{\frac{\omega _{n}}{\hbar }}\right)\exp \left(-\frac{\omega _{n}\chi _{n}^{2}}{2\hbar }\right),\label{eq:psi-nk}\end{equation}
 where $H_{k}\left(x\right)$ are Hermite polynomials and $\omega _{n}\left(a\right)$
is given by Eq.~(\ref{eq:omega-def}). The $\chi _{n}$-dependent
part of the wave function is decomposed into a superposition of instantaneous
excited states,\begin{equation}
\Psi _{n}\left(a,\chi _{n}\right)=\sum _{k=0}^{\infty }C_{k}^{(n)}\left(a\right)\psi _{k}^{(n)}\left(a,\chi _{n}\right),\label{eq:psi-expansion}\end{equation}
 with unknown $a$-dependent coefficients $C_{k}^{(n)}\left(a\right)$.
The complete wave function is a product of these factors,\[
\Psi \left(a,\left\{ \chi _{n}\right\} \right)=e^{-S\left(a\right)}\prod _{n}\Psi _{n}\left(a,\chi _{n}\right).\]

An approximate solution obtained in Ref.~\cite{R84} is of the form
\begin{equation}
C_{2k}^{(n)}\left(a\right)=\frac{\sqrt{\left(2k\right)!}}{k!}\left[\zeta _{n}\left(a\right)\right]^{k},\label{eq:Ck-Rubakov}\end{equation}
 where $\zeta _{n}\left(a\right)$ is a certain explicitly obtained
function, which quickly grows with $a$ when starting at $\zeta _{n}\left(a_{1}\right)=0$,
making the amplitudes $C_{2k}^{(n)}\left(a\right)$ large. However,
the expansion of Eq.~(\ref{eq:psi-expansion}) with the coefficients
from Eq.~(\ref{eq:Ck-Rubakov}) is meaningful only when $\left|\zeta _{n}\left(a\right)\right|<1$.
In Ref.~\cite{R84}, the coefficients $C_{k}^{(n)}$ are interpreted
as amplitudes for the mode $\chi _{n}$ to be in $k$-th excited state,
with the conclusion that such large amplitudes $C_{k}^{(n)}$ demonstrate
a catastrophic particle production and a breakdown of the perturbative
and/or the WKB approximation. Although the coefficients $C_{k}^{(n)}$
should not be interpreted as real particle numbers in the tunneling
regime, it is clear that the backreaction cannot be neglected if $C_{k}^{(n)}$
become large.

The choice of the initial value $\zeta _{n}\left(a_{1}\right)=0$
was motivated in Ref.~\cite{R84} by the intention to investigate
the particle production during tunneling and to start with very few
or no particles in the recollapsing universe. Thus the state with
no particles instantaneously at $a=a_{1}$, \emph{i.e.~}$C_{0}\left(a_{1}\right)=1$,
$C_{2k}\left(a_{1}\right)=0$ for $k\geq 1$, was chosen. However,
the definitions of particles and of the vacuum in an expanding universe
are notoriously ambiguous (see, e.g., \cite{Birrell}), and the method
of instantaneous diagonalization is known to give unphysical results
in some cases \cite{Parker,Fulling}. An unambiguous definition of
particles is possible only in certain very special models. One such
model is a massless conformally coupled scalar field, for which there
is strictly no particle production \cite{Parker}. We shall show in
the companion paper \cite{JVW2} that the state chosen by Rubakov
\emph{et al.~}cannot be considered a vacuum state. In the next section
we shall interpret the states described by Eq.~(\ref{eq:Ck-Rubakov})
as squeezed states.

\subsection{Instantaneously squeezed states}

A squeezed vacuum state $\left|\zeta \right\rangle $ of a harmonic
oscillator can be defined using the creation operator $a^{\dagger }$
and the vacuum state $\left|0\right\rangle $ as\begin{equation}
\left|\zeta \right\rangle =\left(1-\left|\zeta ^{2}\right|\right)^{\frac{1}{4}}\exp \left[-\frac{\zeta }{2}\left(a^{\dagger }\right)^{2}\right]\left|0\right\rangle .\end{equation}
 Here $\zeta $ is the {}``squeezing parameter'', a complex number
satisfying $\left|\zeta \right|<1$. The state is normalized with
the given prefactor, so that $\left\langle \zeta |\zeta \right\rangle =1$.
Since\begin{equation}
\left(a^{\dagger }\right)^{k}\left|0\right\rangle =\sqrt{k!}\left|k\right\rangle ,\label{eq:es1}\end{equation}
 we find that the squeezed state is the following superposition of
the excited states,\begin{equation}
\left|\zeta \right\rangle =\left(1-\left|\zeta ^{2}\right|\right)^{\frac{1}{4}}\sum _{k=0}^{\infty }\frac{\left(-\zeta \right)^{k}\sqrt{\left(2k\right)!}}{2^{k}k!}\left|2k\right\rangle .\label{eq:sqs}\end{equation}
 {[}The normalized wave functions for the excited states are given
by Eq.~(\ref{eq:psi-nk}) and the choice of phases in that equation
is consistent with Eq.~(\ref{eq:es1}).{]}

Suppose that at a particular moment of time the wave function of the
harmonic oscillator with frequency $\omega _{n}$ is a Gaussian, \begin{equation}
\psi _{G}\left(\chi _{n}\right)\propto \exp \left(-\frac{S_{n}}{2}\chi _{n}^{2}\right),\label{eq:wf-chi-n}\end{equation}
 and assume that $Re\, S_{n}>0$. We can represent the wave function
of Eq.~(\ref{eq:wf-chi-n}) by a superposition of excited states,\begin{equation}
\left|\psi _{G}\right\rangle =\sum _{k=0}^{\infty }C_{k}\left|k\right\rangle .\end{equation}
 Only even $k=2p$ will have nonzero amplitudes $C_{k}$. Using the
following formula for the integral of a Hermite polynomial (transformed
from Eq.~22.13.17 of \cite{AS64}),\begin{equation}
\int _{-\infty }^{+\infty }H_{2p}\left(x\right)\exp \left(-\frac{bx^{2}}{2}\right)dx=\sqrt{\pi }\left(2p-1\right)!!\left(2-b\right)^{p}\left(\frac{2}{b}\right)^{p+1/2},\end{equation}
 where $b>0$, we obtain, up to a $p$-independent factor, \begin{equation}
C_{2p}\propto \left(\frac{\omega _{n}-S_{n}}{\omega _{n}+S_{n}}\right)^{p}\frac{\left(2p-1\right)!!}{\sqrt{\left(2p\right)!}}.\end{equation}

Comparing this with Eq.~(\ref{eq:sqs}) and using the identity\begin{equation}
\left(2p-1\right)!!=\frac{\left(2p\right)!}{2^{p}p!},\end{equation}
 we find that the $a$-dependent decomposition of the Gaussian wave
function of Eq.~(\ref{eq:wf-chi-n}) into instantaneous excited states
is exactly the same as that of a squeezed vacuum state with the ($a$-dependent)
squeezing parameter \begin{equation}
\zeta _{n}\left(a\right)=\frac{S_{n}\left(a\right)-\omega _{n}\left(a\right)}{S_{n}\left(a\right)+\omega _{n}\left(a\right)}.\label{eq:ourzeta}\end{equation}
 Incidentally, from Eq.~(\ref{eq:Snequ}) it follows that the function
$\zeta _{n}\left(a\right)$ in the underbarrier region satisfies the
equation\begin{equation}
\frac{d\zeta _{n}}{d\tau }=2\zeta _{n}\omega _{n}-\frac{\dot{\omega }}{2\omega }\left(1-\zeta _{n}^{2}\right).\label{eq:zeta-n-equ}\end{equation}
This equation will be useful in the companion paper \cite{JVW2}.

Thus we have identified the function $\zeta _{n}\left(a\right)$ from
Eq.~(\ref{eq:Ck-Rubakov}) as the instantaneous squeezing parameter.
This allows us to relate the change of the signature of the Gaussian
solution with Rubakov's {}``catastrophic particle production''.
The instantaneous mean occupation number in a squeezed state of Eq.~(\ref{eq:sqs})
is\begin{equation}
\left\langle N\right\rangle =\left(1-\left|\zeta ^{2}\right|\right)^{\frac{1}{2}}\sum _{p=0}^{\infty }\frac{\left|\zeta \right|^{2k}\left(2p\right)!}{2^{2p}\left(p!\right)^{2}}2k=\frac{\left|\zeta \right|^{2}}{1-\left|\zeta \right|^{2}}.\label{eq:sq-N}\end{equation}
 The mean occupation number (in one mode) becomes formally infinite
when $\left|\zeta \right|$ grows above $1$. However, Eq.~(\ref{eq:ourzeta})
can give $\left|\zeta _{n}\left(a\right)\right|\geq 1$ only when
$Re\, S_{n}\left(a\right)\leq 0$. Therefore, the mean occupation
number remains finite, and an apparent catastrophe is avoided, as
long as Eq.~(\ref{eq:Sn-pos}) holds.

We still have not given a physical motivation for the condition of
Eq.~(\ref{eq:Sn-pos}). In the next section we shall address this
issue by considering an exactly solvable example.

\section{Explicit results for a massless field}

In this section we consider the model of a tunneling universe with
a conformally coupled massless scalar field. It is well known that
there is no particle production in this model \cite{Parker}. However,
the arguments of Rubakov \emph{et al.}~would still suggest a {}``catastrophic
particle production'' for some states of the field. To understand
the origin of this discrepancy, it is instructive to compare the Gaussian
solution with the exact wave function obtained by separation of variables
in the Wheeler-DeWitt equation.

\subsection{Solution by separation of variables}

\label{sub:Solution-by-separation}In the massless case, the Wheeler-DeWitt
equation is\begin{equation}
\left[\hbar ^{2}\frac{\partial ^{2}}{\partial a^{2}}-V\left(a\right)-\sum _{n}n^{2}\left(\hbar ^{2}\frac{\partial ^{2}}{\partial \chi _{n}^{2}}-n^{2}\right)\right]\Psi \left(a,\left\{ \chi _{n}\right\} \right)=0.\label{eq:wdw-m0}\end{equation}
 {[}Here we have inserted the degeneracy factor $n^{2}$.{]} The variables
in Eq.~(\ref{eq:wdw-m0}) separate. The separable solutions are of
the form\begin{equation}
\Psi \left(a,\left\{ \chi _{n}\right\} \right)=\psi \left(a\right)\prod _{n}\psi _{n}\left(\chi _{n}\right).\label{eq:psi-sep}\end{equation}
 A general wave function is a linear combination of such solutions.

A single separable solution of the form (\ref{eq:psi-sep}) may be
parametrized by integer occupation numbers $p_{n}\geq 0$, indicating
the excitation levels $p_{n}$ of the modes $\chi _{n}$. The eigenfunctions
in $\chi _{n}$ are given by Eq.~(\ref{eq:psi-nk}) with the substitutions
$\omega _{n}\equiv n$ and $k\equiv p_{n}$. For a given set of the
occupation numbers $\left\{ p_{n}\right\} $, the eigenfunction in
$a$ satisfies\begin{equation}
\left[\hbar ^{2}\frac{\partial ^{2}}{\partial a^{2}}-V\left(a\right)+\hbar \sum _{n}n^{3}\left(2p_{n}+1\right)\right]\psi _{\left\{ p_{n}\right\} }\left(a\right)=0.\end{equation}
 Note that since the number of modes is infinite, the above sum diverges
even if all $p_{n}=0$; this is the divergence of the zero-point vacuum
energy. To obtain a meaningful solution, we assume that the zero-point
energy divergence is absorbed into the radiation density parameter
$\varepsilon _{r}$, and that only finitely many of $p_{n}$ are nonzero.
Then the equation for the eigenfunction $\psi \left(a\right)$ becomes
\begin{equation}
\left[\hbar ^{2}\frac{\partial ^{2}}{\partial a^{2}}-V\left(a\right)+\hbar \sum _{n}2n^{3}p_{n}\right]\psi _{\left\{ p_{n}\right\} }\left(a\right)=0.\label{eq:psia-equ}\end{equation}
 The $2n^{3}p_{n}$ term represents the backreaction of the scalar
field excitations on the background geometry. This term becomes significant
if we consider excited states with large $p_{n}$.

Equation (\ref{eq:psia-equ}) can be solved using the WKB approximation,\begin{equation}
\psi _{\left\{ p_{n}\right\} }\left(a\right)=\exp \left(-\frac{S\left(a\right)}{\hbar }\right),\label{eq:wpn}\end{equation}
 with $S\left(a\right)$ satisfying the equation\begin{equation}
\frac{dS}{da}=\pm \sqrt{V\left(a\right)-\hbar \sum _{n}2n^{3}p_{n}}.\label{eq:S0eq}\end{equation}
 We have kept the $O\left(\hbar \right)$ term in Eq.~(\ref{eq:S0eq})
because we would like to allow arbitrary combinations of excitation
levels, which may lead to large values of $\sum _{n}2n^{3}p_{n}$.

Because of the two possible signs at the square root, we obtain two
branches corresponding to the growing and the decaying solutions under
the barrier, \begin{equation}
S^{g}=\int _{a_{1}}^{a}\sqrt{V\left(a\right)-\hbar \sum _{n}2n^{3}p_{n}}\, da\, ,\quad S^{d}=\int _{a}^{a_{2}}\sqrt{V\left(a\right)-\hbar \sum _{n}2n^{3}p_{n}}\, da\, .\label{eq:S0-gd}\end{equation}
 Here, the boundaries $a_{1,2}$ of the classically forbidden region
$a_{1}<a<a_{2}$ are the two positive roots of the equation \begin{equation}
V\left(a\right)-\hbar \sum _{n}2n^{3}p_{n}=0.\end{equation}

The boundary conditions for the $a$-dependent part of the wave function
are a normalization condition, e.g.~$\psi \left(a=0\right)=1$, and
the tunneling boundary condition at large $a$.

In the case of zero occupation numbers (the vacuum state), the general
(WKB) solution of Eq.~(\ref{eq:wdw-m0}) under the barrier is a linear
combination of the two branches,\begin{equation}
\Psi \left(a,\chi _{n}\right)=\left(C_{g}e^{-\frac{S_{0}^{g}}{\hbar }}+C_{d}e^{-\frac{S_{0}^{d}}{\hbar }}\right)\exp \left[-\frac{1}{2\hbar }\sum _{n}n\chi _{n}^{2}\right].\label{eq:ssol}\end{equation}
 The subscript $0$ in $S_{0}^{d,g}$ signifies that they are the
vacuum solutions given by Eq.~(\ref{eq:S0-gd}) with all $p_{n}=0$.
The relation between the constants $C_{g}$ and $C_{d}$ is to be
obtained from the boundary condition at $a=\infty $ and will not
be important for what follows.

An excited state of the mode $\chi _{n'}$ is given by the wave function\begin{equation}
\Psi \left(a,\chi _{n}\right)\propto H_{p}\left(\chi _{n'}\sqrt{\frac{n'}{\hbar }}\right)\left(C_{g}e^{-\frac{S_{(p)}^{g}}{\hbar }}+C_{d}e^{-\frac{S_{(p)}^{d}}{\hbar }}\right)\exp \left[-\frac{1}{2\hbar }\sum _{n}n\chi _{n}^{2}\right],\label{eq:exc}\end{equation}
 where the functions $S_{(p)}^{g,d}$ are given by Eq.~(\ref{eq:S0-gd})
with all the occupation numbers $p_{n}$ equal to zero except $p_{n'}=p$.
These functions are different from the vacuum functions $S_{0}^{g,d}$
because of the backreaction of the scalar field excitations on the
metric.

Equations (\ref{eq:ssol})-(\ref{eq:exc}) apply in the under-barrier
region, $a_{1}<a<a_{2}$; analogous expressions can be written for
the other regions.

Because of the separation of variables, the definition of the vacuum
and the particle interpretation of the wave function are unambiguous.
A given wave function $\Psi \left(a,\left\{ \chi _{n}\right\} \right)$
is decomposed into separable solutions of the form of Eq.~(\ref{eq:psi-sep}).
Each branch corresponds to a semiclassical universe with a fixed set
of occupation numbers $\left\{ p_{n}\right\} $ and a modified background
geometry. Since the background geometry is affected by the excitations,
a linear superposition of several such semiclassical wave functions
with sufficiently different sets of occupation numbers $\left\{ p_{n}\right\} $
may not correspond to a universe with a definite semiclassical geometry
or particle numbers. However, any classical observers in such a universe
will find themselves in certain semiclassical branches where the spacetime
is fixed and the particle numbers remain constant. In this sense,
there is no observable particle production in the massless model,
for any choice of the quantum state of the universe.

\subsection{Solution in the Gaussian approximation}

If $S_{0}\left(a\right)$ satisfies Eq.~(\ref{eq:S0eq}) with either
of the signs at the square root, then Eq.~(\ref{eq:SnequWKB}) can
be solved in quadratures (for any $V\left(a\right)$). We again introduce
the conformal time variable $\tau $ by Eq.~(\ref{eq:taudef}). The
general solution of Eq.~(\ref{eq:SnequWKB}) with the identification
$\omega _{n}\equiv n$ (as appropriate for the massless case) is\begin{equation}
S_{n}\left(a\right)=n\frac{e^{-2n\tau \left(a\right)}+B}{e^{-2n\tau \left(a\right)}-B},\end{equation}
 where $B$ is an arbitrary constant of integration. We can rewrite
the solutions for the growing and the decaying branches as\begin{equation}
S_{n}^{g}\left(a\right)=n\frac{1+B_{g}e^{2n\tau \left(a\right)}}{1-B_{g}e^{2n\tau \left(a\right)}},\quad S_{n}^{d}\left(a\right)=n\frac{1+B_{d}e^{-2n\tau \left(a\right)}}{1-B_{d}e^{-2n\tau \left(a\right)}},\label{eq:esol}\end{equation}
 where we have introduced the constants $B_{g}$ and $B_{d}$ for
these two branches. (We suppressed the index $n$ at the constants
$B_{g}$, $B_{d}$ for brevity.)

The choice of the constants $B_{g,d}$ is determined by the initial
condition $S_{n}\left(a_{1}\right)$ and by matching of the branches
at the second turning point $a_{2}$. It is clear from Eq.~(\ref{eq:esol})
that some choices of the constants will lead to negative values of
$S_{n}\left(a\right)$ under the barrier. A special choice is $B_{g}=B_{d}=0$;
in this case, the function $S_{n}\left(a\right)$ is constant, and
the vacuum solution of Eq.~(\ref{eq:ssol}) is recovered. For other
values of $B$, the Gaussian solution corresponds to a squeezed state.
The corresponding squeezing parameter at $a=a_{1}$ is $\zeta _{n}\left(a_{1}\right)=B$,
as follows from Eq.~(\ref{eq:ourzeta}) with the identification $\omega _{n}\equiv n$
appropriate for the massless case.

The squeezing parameter $\zeta _{n}\left(a\right)$ satisfies Eq.~(\ref{eq:zeta-n-equ})
where $\omega _{n}=const$, \begin{equation}
\frac{d\zeta _{n}}{d\tau }=2n\zeta _{n}.\end{equation}
 For a given initial squeezing parameter $\zeta _{n}\left(a_{1}\right)$,
we find that, at a point $a=a_{*}$ under the barrier, $\left|\zeta _{n}\left(a_{*}\right)\right|=1$
and therefore $Re\, S_{n}\left(a_{*}\right)=0$ if \begin{equation}
\left|\zeta _{n}\left(a_{1}\right)\right|=\exp \left[-2n\tau \left(a_{*}\right)\right].\label{eq:zeta-cond1}\end{equation}
 Therefore, Eq.~(\ref{eq:Sn-pos}) will hold everywhere under the
barrier ($a_{1}<a<a_{2}$) only if $\zeta _{n}\left(a_{1}\right)$
is small enough so that\begin{equation}
\left|\zeta _{n}\left(a_{1}\right)\right|<\exp \left[-2n\tau \left(a_{2}\right)\right].\label{eq:zeta-limit}\end{equation}

The special choice $B=0$ corresponds to the (unique) vacuum state
$\zeta _{n}\equiv 0$. With this choice, the Gaussian solution will
always be well-behaved, with $Re\, S_{n}>0$ everywhere.

\subsection{Comparison: The case of $H=0$}

\label{sub:Comparison}

If $H=0$, the barrier is infinitely high, so there is no tunneling
but only the recollapsing universe. We use this case to illustrate
the underbarrier behavior of the wave function, because explicit calculations
can be more easily done. In the $H\neq 0$ case, the potential $V\left(a\right)$
is well approximated by the $H=0$ expression, $V\left(a\right)=a^{2}-\varepsilon _{r}$,
when $a$ is sufficiently small, $a\ll H^{-1}$.

\begin{figure}[htbp]
\begin{center}\includegraphics[  bb=72bp 72bp 396bp 306bp,
  width=4in,
  keepaspectratio]{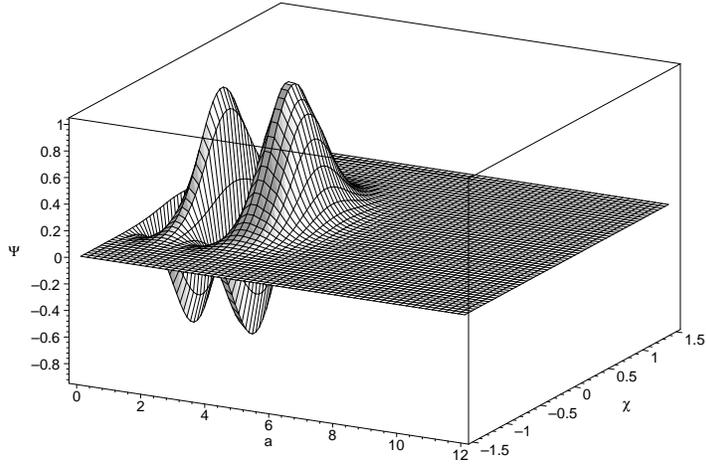}\end{center}

\caption{The vacuum wave function ($\zeta =0$) in the $a-\chi $ space with
$\varepsilon _{r}=15$. (Only the mode $\chi _{n}$ with $n=10$ is
shown, and other modes $\chi _{n}$ are omitted.) Oscillatory behavior
until the turning point $a_{1}\approx 4$ is followed by an exponential
decay under the barrier ($a>a_{1}$).\label{fig:wf0}}
\end{figure}

The Gaussian solution for $H=0$, together with the condition $Re\, S_{n}\left(a\right)>0$
for all $a$, gives a unique wave function, because Eq.~(\ref{eq:zeta-limit})
forces $\zeta =0$. The Gaussian wave function with $\zeta =0$ corresponds
to the vacuum state {[}given by Eq.~(\ref{eq:ssol}){]} and is illustrated
in Fig.~\ref{fig:wf0}.%

We now consider the wave function for a squeezed state with $\zeta \equiv \zeta \left(a_{1}\right)\ne 0$.
We can construct the exact wave function using Eqs.~(\ref{eq:sqs}),
(\ref{eq:exc}) and compare it with the Gaussian solution. %
\begin{figure}[htbp]
\begin{center}\includegraphics[  bb=72bp 72bp 396bp 306bp,
  width=4in,
  keepaspectratio]{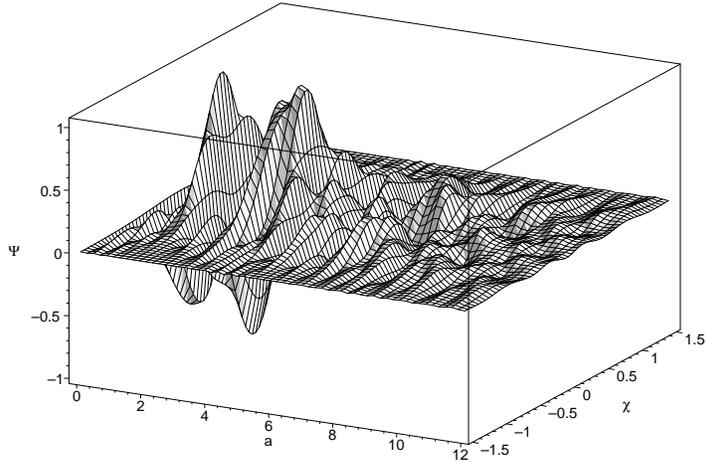}\end{center}

\caption{The wave function of a squeezed state ($\zeta =0.5$) in the $a-\chi $
space, with the same parameters as in Fig.~\ref{fig:wf0}. The underbarrier
($a>a_{1}$) wave function is dominated by highly excited states.\label{fig:wf1}}
\end{figure}

An explicit calculation of the exact wave function $\Psi \left(a,\chi _{n}\right)$
is given in Appendix A. At $a=a_{1}$ it has the form of a squeezed
state with a squeezing parameter $\zeta \equiv \zeta \left(a_{1}\right)$.
For $\zeta $ not too close to $1$, the wave function is dominated
by the vacuum state and nearby states with low occupation numbers.
However, as we go under the barrier, the contribution of highly excited
states becomes increasingly important, and at large enough $a$ they
completely dominate the wave function. This happens when $a$ exceeds
the following $\zeta $-dependent bound, \begin{equation}
a>a_{\max }\left(\zeta \right)\equiv \frac{\sqrt{\varepsilon _{r}}}{2}\left(\zeta ^{-\frac{1}{2n}}+\zeta ^{\frac{1}{2n}}\right).\label{eq:amax}\end{equation}
 These excited states belong to different semiclassical branches of
the wave function, and thus the picture of a quantum field in a semiclassical
background spacetime does not apply. The wave functions of the excited
states exhibit short-wavelength oscillations in the $\chi $-direction,
which are clearly visible in Fig.~\ref{fig:wf1}. This signals the
breakdown of the Gaussian approximation. Now, it can be easily verified
that the value $a_{\max }$ in Eq.~(\ref{eq:amax}) coincides with
$a_{*}$ defined in Eq.~(\ref{eq:zeta-cond1}) as the value of $a$
where $S_{n}\left(a\right)$ changes sign. This follows from (\ref{eq:zeta-cond1})
after substitution of the expression \begin{equation}
\tau \left(a\right)=\cosh ^{-1}\frac{a}{\sqrt{\varepsilon _{r}}}\end{equation}
 for the Euclidean conformal time for $H=0$. If $\zeta \neq 0$ then
$a_{\max }$ is finite and $Re\, S_{n}\left(a\right)$ becomes negative
for $a>a_{\max }$. Only if $\zeta =0$, the condition of Eq.~(\ref{eq:Sn-pos})
holds for all $a>0$. We find that Eq.~(\ref{eq:Sn-pos}) is indeed
the condition of consistency of the Gaussian approximation.

We wish to stress that the constant parameter $\zeta \equiv \zeta (a_{1})$
employed in the Appendix is not to be confused with the instantaneous
squeezing parameter $\zeta \left(a\right)$ defined by Eq.~(\ref{eq:ourzeta}).
The instantaneous squeezing parameter $\zeta \left(a\right)$ does
not necessarily reflect the true particle content of the quantum state.
In particular, $\left|\zeta \left(a\right)\right|>1$ when $Re\, S_{n}\left(a\right)<0$,
and a squeezed state with $\left|\zeta \right|>1$ is ill-defined
and formally resembles a state with infinitely many particles, $\left\langle N\right\rangle =\infty $.
The interpretation of Ref.~\cite{R84} would suggest a {}``catastrophic
particle production'' due to tunneling if the barrier is wide enough
so that $a_{2}>a_{\max }$. But, in fact, the model with a massless
conformally coupled scalar field does not have any particle production.

A physical explanation of this result has already been partially given
in Ref.~\cite{R84}. A squeezed state is a superposition of all excited
states, and the amplitudes of high-energy excited states are exponentially
suppressed. On the other hand, the low-energy states contribute an
exponentially suppressed amount to the wave function because of tunneling.
But the two exponential suppression factors have a different behavior
because the exponential suppression of low-energy states is $a$-dependent.
Therefore, the high-energy states (which were already contained in
the recollapsing universe) will give a dominant contribution to the
wave function at large enough $a$ under the barrier.

\section{Discussion }

We have examined the solutions of the Wheeler-DeWitt equation for
the FRW universe with a conformally coupled massless scalar field.
We have shown that the wave function under the barrier exhibits all
the signs of a {}``catastrophic particle production'' in the sense
of Rubakov \emph{et al.}, even though there is no actual particle
production in this model. The calculations of Rubakov \emph{et al.}~are
formally correct, but we disagree with their interpretation.

The wave function of the universe is a superposition of different
semiclassical geometries; each semiclassical universe contains a different
quantum state of the scalar field. If the quantum state of the universe
is a superposition of low-energy and high-energy states of the scalar
field, then the wave function of the expanding universe will be dominated
by the semiclassical geometries that essentially did not tunnel, rather
than by the geometry of an inflating universe created by tunneling.

For example, in the massless model we might consider a quantum state
of the universe which is a superposition of the vacuum state $\Psi _{0}$
and of the state $\Psi _{n}^{k}$ which is the $k$-th excited state
of a single mode $\chi _{n}$ of the quantum field. A superposition
of these states such as $\Psi _{0}+\alpha \Psi _{n}^{k}$, with a
small amplitude $\alpha $, will be close to the vacuum state to the
left of the barrier (in the recollapsing universe). However, to the
right of the barrier the contribution of the state $\Psi _{n}^{k}$
will dominate the wave function, and the inflating universe will appear
to be in an excited state with the occupation number $k$ in the mode
$\chi _{n}$.

This result should be interpreted not as a production of particles
during tunneling, but rather as an emergence of excited states that
have been already present in the recollapsing universe and became
dominant in the expanding regime after tunneling. Had these highly
excited states not been present, the final state would have been that
of an empty inflating universe. Thus, to investigate the creation
of the universe through quantum tunneling, the initial state of the
recollapsing universe must be chosen correctly.

In the model with a conformally coupled massless scalar field, there
is a preferred choice of the quantum state of the recollapsing universe
which does not contain any admixture of excited states. This quantum
state can be identified with the vacuum state. We have performed an
explicit calculation to demonstrate that the wave function of any
(non-vacuum) squeezed state becomes dominated by high-energy states
exactly at the same region where the perturbative formalism of Rubakov
\emph{et al.}~starts to break down.

We have also shown that the Gaussian solution of the Wheeler-DeWitt
equation {[}Eq.~(\ref{eq:wfansatz}){]} is equivalent to a squeezed
vacuum state in the formalism of Rubakov \emph{et al}. The Gaussian
solution also manifests the domination by high-energy states, in that
the wave function starts to grow at large $\chi $. Therefore, the
condition (\ref{eq:Sn-pos}) that the Gaussian wave function decreases
at large $\chi $ can be used to select a quantum state of the universe
which is dominated by the vacuum state rather than by the admixture
of high-energy excited states.

In the companion paper \cite{JVW2}, we shall extend our conclusions
to the more general case of a massive conformally coupled field where,
unlike the case of the massless field, one would expect some particle
creation. We shall demonstrate that the quantum state of the recollapsing
universe can be chosen to contain a sufficiently small admixture of
excited states, and that in this case the WKB approximation is everywhere
applicable and the backreaction of the matter excitations on the metric
is negligible. We note that the under-barrier behavior of a massive
field has been discussed by Bouhmadi-López, Garay and González-Díaz
\cite{BGG02}, who constructed a vacuum wave function satisfying the
condition of Eq.~(\ref{eq:Sn-pos}) in the case of a negative cosmological
constant ($H^{2}<0$). The paper \cite{BGG02} has a significant overlap
with our work in \cite{JVW2}, and we shall comment on it there in
more detail.

\section*{Acknowledgments}

We are grateful to Larry Ford, Jaume Garriga, Xavier Siemens, and
Takahiro Tanaka for stimulating discussions. This work was supported
in part by a NSF grant.

\appendix

\section{The under-barrier wave function of a squeezed state}

Here we compute the wave function of a squeezed state with an arbitrary
squeezing parameter $\zeta \neq 0$ in the case $H=0$. The potential
$V\left(a\right)$ is \begin{equation}
V\left(a\right)=a^{2}-\varepsilon _{r}.\label{eq:va0}\end{equation}
 For simplicity we consider the vacuum state in all modes except one
particular mode $\chi _{n}$. {[}Simultaneous squeezed states of several
modes give analogous results.{]} Our purpose is to show that the wave
function under the barrier is dominated by the contribution of certain
high-energy excited states, rather than by the vacuum solution, and
to find the relevant range of $a$.

For an excited state of the mode $\chi _{n}$ with occupation number
$2k$, the turning point is at\begin{equation}
a_{0}\left(2k\right)=\sqrt{\varepsilon _{r}+4n\hbar k}.\end{equation}
 For a given $a$, excited states with $k>k_{\min }\left(a\right)$
are above the barrier, where\begin{equation}
k_{\min }=\frac{V\left(a\right)}{4n\hbar }.\end{equation}
The wave function for an excited state of level $2k$ is {[}cf.~Eq.~(\ref{eq:psi-nk}){]}\begin{equation}
\Psi _{2k}\left(a,\chi _{n}\right)=\left(\frac{n}{\pi \hbar }\right)^{\frac{1}{4}}\frac{H_{2k}\left(\chi _{n}\sqrt{\frac{n}{\hbar }}\right)}{\sqrt{\left(4k\right)!!}}\exp \left(-\frac{n\chi _{n}^{2}}{2\hbar }\right)\psi _{2k}\left(a\right),\label{eq:psi2k}\end{equation}
 where the $a$-dependent part $\psi _{2k}\left(a\right)$ is given
by a WKB approximation,\begin{equation}
\psi _{2k}\left(a\right)\propto \left\{ \begin{array}{l}
 \frac{1}{\sqrt[4]{V\left(a\right)-4n\hbar k}}\exp \left(-\frac{1}{\hbar }\int _{a_{0}\left(2k\right)}^{a}\sqrt{V\left(a\right)-4n\hbar k}\, da\right),\quad a>a_{0}\left(2k\right),\\
 \frac{2}{\sqrt[4]{4n\hbar k-V\left(a\right)}}\cos \left(\frac{1}{\hbar }\int _{a}^{a_{0}\left(2k\right)}\sqrt{4n\hbar k-V\left(a\right)}\, da-\frac{\pi }{4}\right),\quad 0<a<a_{0}\left(2k\right).\end{array}
\right.\end{equation}

The wave function $\Psi \left(a,\left\{ \chi _{l}\right\} \right)$
is a superposition of the wave functions for excited states of the
mode $\chi _{n}$ (we suppress the dependence on other modes $\chi _{l}$
with $l\neq n$), with coefficients given by Eq. (\ref{eq:sqs}) with
the squeezing parameter $\zeta $,\begin{eqnarray}
\Psi \left(a,\chi _{n}\right) & = & \left(1-\left|\zeta \right|^{2}\right)^{\frac{1}{4}}\left(\frac{n}{\pi \hbar }\right)^{\frac{1}{4}}\exp \left(-\frac{n\chi _{n}^{2}}{2\hbar }\right)\nonumber \\
 &  & \times \sum _{k=0}^{\infty }\frac{\left(-\zeta \right)^{k}\sqrt{\left(2k\right)!}}{2^{k}k!}\frac{H_{2k}\left(\chi _{n}\sqrt{\frac{n}{\hbar }}\right)}{\sqrt{\left(4k\right)!!}}\psi _{2k}\left(a\right).\label{eq:sum0}
\end{eqnarray}

For the analysis below we will need an asymptotic formula for Hermite
polynomials $H_{n}\left(x\right)$ at fixed $x$ and large $n$,\begin{equation}
H_{n}\left(x\right)\sim \sqrt{2}\left(2n\right)^{\frac{n}{2}}e^{\frac{x^{2}}{2}-\frac{n}{2}+O(n^{-1})}\cos \left(-\frac{\pi n}{2}+x\sqrt{2n}+O(n^{-1/2})\right).\label{eq:Hn}\end{equation}
 This expression can be derived by the method of steepest descent
from the integral representation (cf.~\cite{AS64}, Eq.~22.10.15)\begin{equation}
H_{n}\left(x\right)=\frac{i^{n}}{2\sqrt{\pi }}\int _{-\infty }^{+\infty }e^{-\left(\frac{t}{2}+ix\right)^{2}}t^{n}dt.\end{equation}

Using the Stirling formula for large factorials\begin{equation}
n!\sim \sqrt{2\pi n}\left(\frac{n}{e}\right)^{n}(1+O\left(n^{-1}\right)),\end{equation}
 we find \begin{equation}
\frac{\sqrt{\left(2k\right)!}}{2^{k}k!}\sim k^{-1/4}.\end{equation}
 Compared with the exponential functions of $k$ in Eq.~(\ref{eq:sum0}),
this is a slowly changing factor.

Consider the contribution of the state $\left|2k\right\rangle $ to
the wave function, as a function of $k$. The contribution of levels
$k>k_{\min }$ decreases with $k$ because of the suppression factor
$\zeta ^{k}$. The absolute value of the contribution of a level $k<k_{\min }$
can be estimated as\begin{equation}
\left|\zeta ^{k}\Psi _{2k}\left(a,\chi _{n}\right)\right|\sim \exp \left(-zk-\frac{1}{\hbar }\int _{a_{0}\left(2k\right)}^{a}\sqrt{V\left(a\right)-4n\hbar k}\, da\right).\label{eq:kcontrib}\end{equation}
 Here we have defined for convenience\begin{equation}
z\equiv -\ln \left|\zeta \right|.\end{equation}
 {[}We have used the asymptotic Eq.~(\ref{eq:Hn}) which is justified
if $k$ is large. There is no dependence on $\chi _{n}$ in the absolute
value of the wave function.{]} For $V\left(a\right)$ given by Eq.~(\ref{eq:va0}),
the integral in Eq.~(\ref{eq:kcontrib}) can be evaluated as\begin{equation}
\int _{a_{0}}^{a}\sqrt{a^{2}-a_{0}^{2}}\, da=\frac{a}{2}\sqrt{a^{2}-a_{0}^{2}}-\frac{a_{0}^{2}}{2}\cosh ^{-1}\frac{a}{a_{0}}.\end{equation}
 The function under the exponential in Eq.~(\ref{eq:kcontrib}) has
a maximum at\begin{equation}
k=k_{0}\left(a,\zeta \right)=\frac{1}{4n\hbar }\left(\frac{a^{2}}{\cosh ^{2}\frac{z}{2n}}-\varepsilon _{r}\right).\end{equation}
 {[}Note that $k_{0}>0$ only for large enough $a$.{]} Therefore
the dominant contribution to the wave function comes from either $k=0$
(for $0<a<\sqrt{\varepsilon _{r}}\cosh \frac{z}{2n}$) or from $k\approx k_{0}\left(a,\zeta \right)$
(for $a>\sqrt{\varepsilon _{r}}\cosh \frac{z}{2n}$). One can see
this in Fig.~\ref{fig:wf1}: at progressively larger values of $a$,
the wave function exhibits more oscillations in the $\chi _{n}$ direction,
which corresponds to excited states with different values of $k=k_{0}\left(a,\zeta \right)$.

We find that the wave function $\Psi \left(a,\chi _{n}\right)$ is
dominated by the contribution of excited states $\Psi _{2k}$ with
$k\approx k_{0}\left(a,\zeta \right)$ when\begin{equation}
a>a_{\max }\equiv \sqrt{\varepsilon _{r}}\cosh \frac{z}{2n}=\frac{\sqrt{\varepsilon _{r}}}{2}\left(\zeta ^{-\frac{1}{2n}}+\zeta ^{\frac{1}{2n}}\right).\end{equation}
 This is the expression we needed in Sec.~\ref{sub:Comparison}.


\begin{thebibliography}{10}
\bibitem{AV86}A. Vilenkin, Phys. Rev. D \textbf{30}, 509 (1984); \textbf{33}, 3560
(1986).
\bibitem{R84}V. A. Rubakov, Phys. Lett. B \textbf{148}, 280 (1984).
\bibitem{ZelStar}Y. B. Zeldovich and A. A. Starobinsky, Sov. Astron. Lett. \textbf{10},
135 (1984).
\bibitem{HH}J. B. Hartle and S. W. Hawking, Phys. Rev. D \textbf{28}, 2960 (1983).
\bibitem{Linde84}A. D. Linde, Lett. Nuovo Cim. \textbf{39}, 401 (1984).
\bibitem{AV02}A. Vilenkin, preprint gr-qc/0204061 (2002).
\bibitem{LRR02}D. Levkov, C. Rebbi, and V. A. Rubakov, preprint gr-qc/0206028 (2002).
\bibitem{V88}A. Vilenkin, Phys. Rev. D \textbf{37}, 888 (1988).
\bibitem{VV88}T. Vachaspati and A. Vilenkin, Phys. Rev. D \textbf{37}, 898 (1988).
\bibitem{GV97}J. Garriga and A. Vilenkin, Phys. Rev. D \textbf{56}, 2464 (1997).
\bibitem{Parker}L. Parker, Phys. Rev. \textbf{183}, 1057 (1969).
\bibitem{JVW2}J. Hong, A. Vilenkin, and S. Winitzki, in preparation.
\bibitem{HH85}J. J. Halliwell and S. W. Hawking, Phys. Rev. D \textbf{31}, 1777
(1985).
\bibitem{LR79}V. Lapchinsky and V. A. Rubakov, Acta Phys. Pol. B \textbf{10}, 1041
(1979).
\bibitem{Banks}T. Banks, Nucl. Phys. B \textbf{249}, 332 (1985).
\bibitem{BBW73}T. Banks, C. Bender, and T. T. Wu, Phys. Rev. D \textbf{8}, 3346 (1973);
\textbf{8}, 3366 (1973).
\bibitem{Birrell}N. D. Birrell and P. C. W. Davies, \textit{Quantum fields in curved
space}, Cambridge University Press, Cambridge, 1982.
\bibitem{Fulling}S. A. Fulling, Gen. Rel. Grav. \textbf{10}, 807 (1979).
\bibitem{AS64}M. Abramowitz and I. A. Stegun, \emph{Handbook of Mathematical Functions},
National Bureau of Standards, 1964.
\bibitem{BGG02}M. Bouhmadi-López, L. J. Garay, and P. F. González-Díaz, preprint
astro-ph/0204072 (2002).\end{thebibliography}
\end{document}